\documentclass[aps, prl, twocolumn, showkeys]{revtex4-1}

\usepackage{chemformula} 
\usepackage{siunitx} 
\usepackage[inline,shortlabels]{enumitem} 
\usepackage{hyperref} 
\usepackage{graphicx} 

\usepackage{natbib}

\begin{document}

    \title{Dopant incorporation site in sodium cobaltate's host lattice: A critical factor for thermoelectric performance}
    \author{M. H. N. Assadi}
    \email{Assadi@aquarius.mp.es.osaka-u.ac.jp}
    \altaffiliation[Tel:]{+81668506671}
    \author{H. K tayama-Yoshida}
    \affiliation{Department of Materials Physics, Graduate School of Engineering Science, Osaka $U$niversity, Osaka 560-8531, Japan}
    \date{2015}

    \begin{abstract}
        \ch{Na_{x}CoO2} that comprises of alternating Na and \ch{CoO2} layers has exotic magnetic and thermoelectric properties that could favorably be manipulated by adding dopants or varying Na concentration. In this work, we investigated the structural and electronic properties of Sr and Sb doped \ch{Na_{x}CoO2} ($x = 0.50, 0.625, 0.75$ and 0.875) through comprehensive density functional calculations. We found that Sr dopants always occupy a site in the Na layer while Sb dopants always substitute a Co ion in the host lattice regardless of Na concentration. This conclusion withstood when either generalized gradient approximation (GGA) or GGA+$U$ method was used. By residing on the Na layer, Sr dopants create charge and mass inertia against the liquid like Na layer, therefore, improving the crystallinity and decreasing the electrical resistivity through better carrier mobility. On the other hand, by substituting Co ions, Sb dopants reduce the electrical conductivity and therefore decrease the Seebeck coefficient.
    \end{abstract}
    \maketitle

    \section{Introduction}
        Layered sodium cobaltate (\ch{Na_{x}CoO2}) is a promising material for high-efficiency thermoelectric energy harvesting at high temperatures.
        The advantage of \ch{Na_{x}CoO2} is in its complex lattice structure that consists of alternating Na layers and edge-sharing \ch{CoO2} octahedral layers \cite{Fergus2012}. The triangular nature of \ch{CoO2} layer creates high electronic frustration that results in large spin entropy \cite{Koshibae2001} and therefore large Seebeck coefficient \cite{Terasaki1997}.
        Additionally, in \ch{Na_{x}CoO2}, heat carrying phonons are strongly scattered by the  \ch{Na+} ions which have a liquid like pattern at temperatures above ambient \cite{Foo2004}.
        This combination offers pathways for favorably and independently adjusting all otherwise interdependent factors of the \textit{figure of merit} ($ZT$) \cite{Mahan1996, Snyder2008}.
        From a materials engineering viewpoint, doping various elements has been extensively explored to improve the \ch{Na_{x}CoO2}'s thermoelectric performance.
        For example, heavier ions such as rare earth elements \cite{Nagira2004} and Ag \cite{Seetawan2006} were used to decrease the lattice thermal conductivity ($k_l$) as a strategy to improve the $ZT$ of the \ch{Na_{x}CoO2} systems.
        In one instance, $5\%$ of Yb dopants significantly increased the Seebeck coefficient to $250\si{.\mu VK^{-1}}$ at $T = 1000\si{.K}$ ($\sim 150\si{.\mu VK^{-1}}$ for pristine \ch{Na_{x}CoO2}), however, with the side effect of increased resistivity \cite{Nagira2004}.
        Due to similar trade-offs in ceramic sodium cobaltate systems doped with various elements, the upper limit for $ZT$ has stagnated at values below one despite the intensive research efforts in recent years \cite{Terasaki2013}.

        Further progress in the field requires a fundamental atomistic level understanding of how the property trade-offs are correlated with the behavior of dopants and Na concentration in \ch{Na_{x}CoO2}.
        One particular concern is that due to the difference in the bonding nature and local chemical environment of the \ch{CoO2} and Na layers in \ch{Na_{x}CoO2} systems, a particular dopant may be stable at different lattice sites as Na concentration varies.
        Interactions governing dopants' behaviors are strongly dominated by the chemical coordination, which can be determined by techniques such as the analysis of x-ray absorption near edge structure analysis. However, such measurements are fundamentally limited by the energy range of the instruments.
        As a result, the theoretical investigation becomes an indispensable tool to provide insight in this regard.
        Theoretically, research on the effect of Na concentration on the electrical and structural properties has been conducted using standard density functional theory (DFT) \cite{Luo2004, Zhang2005, Meng2008, Lee2006a}, DFT plus Monte Carlo approach \cite{Wang2007} and DFT plus Gutzwiller method \cite{Wang2008}.
        However, these studies have mainly been focused on the pristine \ch{Na_{x}CoO2} system only.
        In this work, we investigate the behavior of Sr and Sb dopants in \ch{Na_{x}CoO2} for values of $x = 0.50, 0.625, 0.75$ and 0.875 by density functional theory.
        \ch{Na_{x}CoO2} with higher Na concentration of $x > 0.5$, as investigated here, has excessively higher thermopower and thus is appealing for practical applications \cite{Lee2006b}.
        We chose Sr and Sb as a dopant particularly because they have exceedingly higher atomic masses than all other elements of the host material thus may favorably affect thermal conductivity.

    \section{Computational Settings}
        All-electron \textit{ab initio} spin-polarized density functional calculations were performed using Accelrys's \ch{DMol^3} package \cite{Delley1990, Delley2000}.
        Self-consistent energy calculations were performed with “double-numeric plus polarization” (DNP) basis set for all electrons while generalized gradient approximation (GGA) based on Perdew-Wang formalism \cite{Perdew1992} was applied for exchange-correlation functional.
        Real-space global cutoff radii were set for all elements at $5.2\si{.\angstrom}$ to warrant accurate numerical integration for Na orbitals.
        Brillouin zone sampling was carried out by choosing a $4\times2\times2$ k--point set within Monkhorst-Park scheme with a grid spacing of $\sim0.05\si{.\angstrom^{-1}}$ between $k$ points.
        Total energy convergence test was performed by increasing the $k$ point mesh to $6\times3\times3$; it was found that the total energy changed by only $10^{-5}\si{.eV/atom}$, therefore the results were well converged.
        Finally, we used “direct inversion in the iterative subspace” algorithm \cite{Csaszar1984} to optimize the ionic position where we set the convergence thresholds for energy, Cartesian components of internal forces and displacement at $10^{-5}\si{.eV/atom}, 0.01\si{.eV/\angstrom}$, and $0.005\si{.\angstrom}$ respectively.
        The lattice parameters of the primitive \ch{NaCoO2} unit cell were found to be $2.87\si{.\angstrom}$ for $a$ and $10.90\si{.\angstrom}$ for $c$ which are in reasonable agreement with experimental lattice parameters \cite{Chen2004} differing by $1.04\%$ and $0.91\%$ for $a$ and $c$ respectively.
        Then a $4a\times2a\times1c$ supercell of \ch{Na16Co16O32} constructed to create the Na deficient systems.
        To obtain the final atomic geometries, the internal coordinates of all ions in the supercell were relaxed while fixing the lattice parameters to their theoretical values of the pristine \ch{NaCoO2} in order to avoid artificial hydrostatic pressure.
        To vary sodium concentration, Na ions were systematically removed from the original supercell to create \ch{Na_{x}CoO2}.
        In Na deficient systems, the lattice parameters vary slightly with Na concentration, however, DFT results are not sensitive to this small variation \cite{Zhang2005}.
        The ground state (lowest energy) arrangements of the Na ions in \ch{Na_{x}CoO2} were adopted after our previous work \cite{Assadi2014}.

        \begin{figure}
            \centering
            \includegraphics[width=0.9\columnwidth]{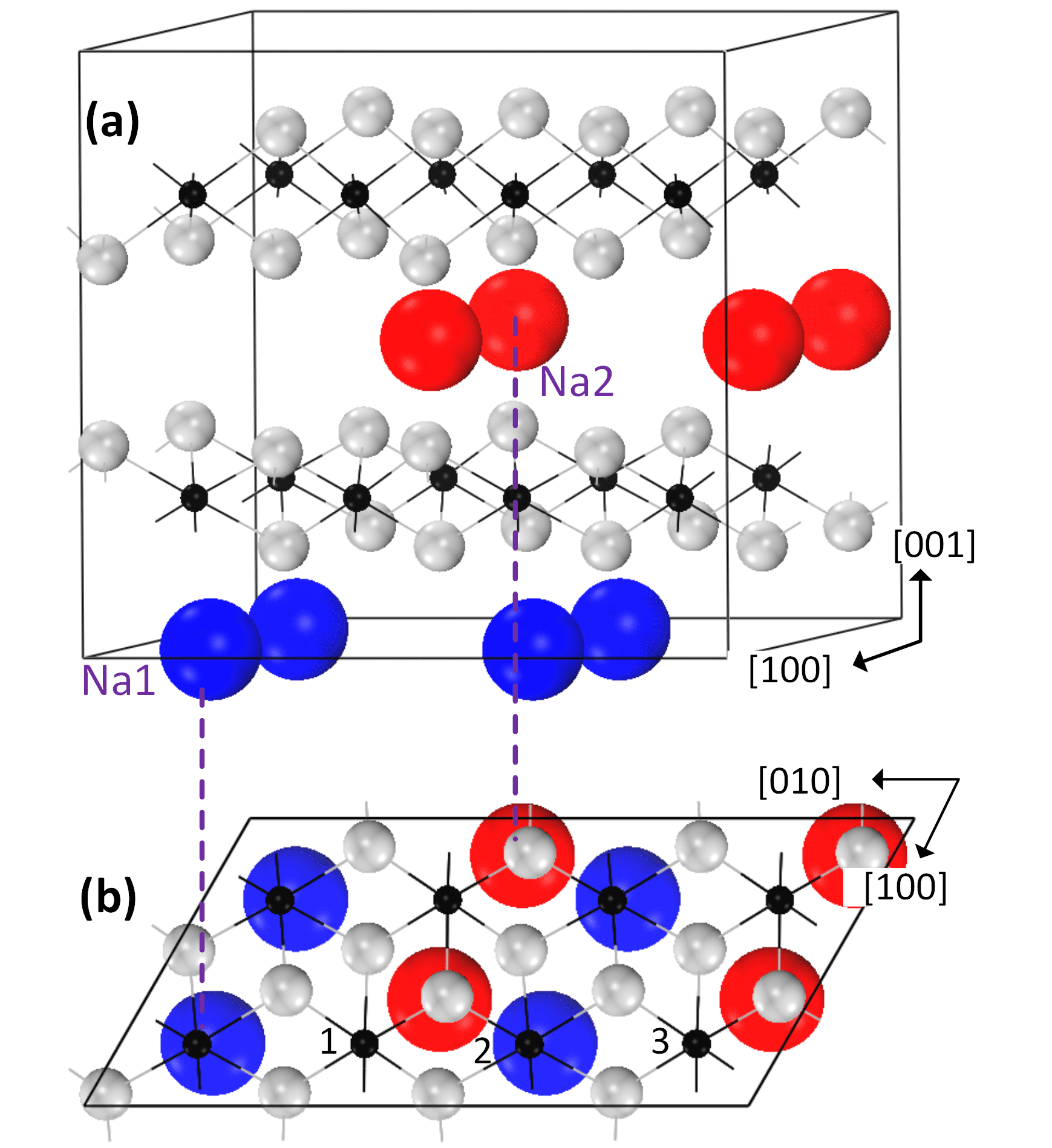}
            \caption{\label{fig:1}Side view (a) and top view (b) of the $4a\times2a\times1c$ supercell of \ch{Na_{0.5}CoO2} is schematically presented as an example of \ch{Na_{x}CoO2} structure. There are two crystallographically distinct sites for Na; Na1 where Na shares basal coordinates with Co with Wyckoff position $b$ and Na2 where Na shares basal coordinates with O with Wyckoff position $d$ within the P6\textsubscript{3} \ch{Na_{x}CoO2} primitive unit cell while Co and O occupy $a$ and $f$ Wyckoff positions respectively. Blue and red spheres represent Na ions with Z = 0 and Z = 0.5 respectively. Black spheres denote Co ions while grey spheres denote O ions.}
        \end{figure}

    \section{Results and Discussions}
        \subsection{Formation Energies of the Sr and Sb Dopants}
            Since we only considered cationic dopants, dopants' formation energy ($E^f$) was calculated for four possible crystallographical configurations for any given Na concentration.
            In the first configuration, the dopant substituted a Na ion at Na1 site creating a \ch{Sr_{Na 1}} or \ch{Sb_{Na 1}} configurations.
            In the second configuration, the dopants substituted a Na ion at Na2 site creating a \ch{Sr_{Na 2}} or \ch{Sb_{Na 2}} configurations.
            Na1 and Na2 are two crystallographically distinct positions of Na ions within the unit cell.
            Na1 shares the same basal coordinates with an Co while Na2 shares the same basal coordinates with a O as demonstrated in Fig. \ref{fig:1}.
            In the third configuration, the dopant occupied an interstitial site in Na layer creating \ch{Sr_{Int}} or \ch{Sb_{Int}} configurations.
            Finally, the fourth configuration was constructed by substituting the dopant for a Co ion creating \ch{Sr_{Co}} or \ch{Sb_{Co}} configurations.
            Since the ionic volume of both Sr and Sb were in the order of $\sim 2 \times 10^{-4}\si{.nm^3}$ which is much larger than the interstitial cavity in \ch{CoO2} layer of $1.79 \times 10^{-7} \si{.nm^3}$, this interstitial site was not considered.
            The formation energy ($E^f$) of the dopants was calculated with the standard procedure \cite{vandeWalle2004} as described by the following equation:
            \begin{equation}
                E^f = E^t \left(\ch{Na_x CoO2} : \ch{X}\right) + \mu_{\alpha} - E^t \left(\ch{Na_x CoO2}\right) - \mu_X.
            \end{equation}
            Here, $E^t \left(\ch{Na_x CoO2} : X\right)$ is the total energy of the \ch{Na_{x}CoO2} supercell containing the dopant X, and $E^t \left(\ch{Na_x CoO2}\right)$  is the total energy of the pristine \ch{Na_{x}CoO2} supercell.
            $\mu_\alpha$ and $\mu_X$ are the chemical potentials of the removed and added elements respectively.
            The chemical potentials of Co, Na, Sb and Sr were calculated from the total energies of their respective most stable oxides that is \ch{CoO}, \ch{NaO2} and \ch{Sb3O4} and \ch{SrO}, representing an oxygen-rich condition, by the following equation:
            \begin{equation}
                \mu = \left.\left[E^t \left(\ch{X}_n \ch{O}_m\right) - \frac{1}{2} m E^t \left(\ch{O2}\right)\right]\right/n
            \end{equation}
            in which $\ch{X}_n \ch{O}_m$ is the most stable oxide of the corresponding elements X (X = Co, Na, Sr and Sb).
            We found that $E^t \left(\ch{O2}\right)$ was $-5.94 \si{.eV}$ which was in good agreement with previous DFT calculations \cite{Behler2004}.

            \begin{figure}
                \centering
                \includegraphics[width=0.9\columnwidth]{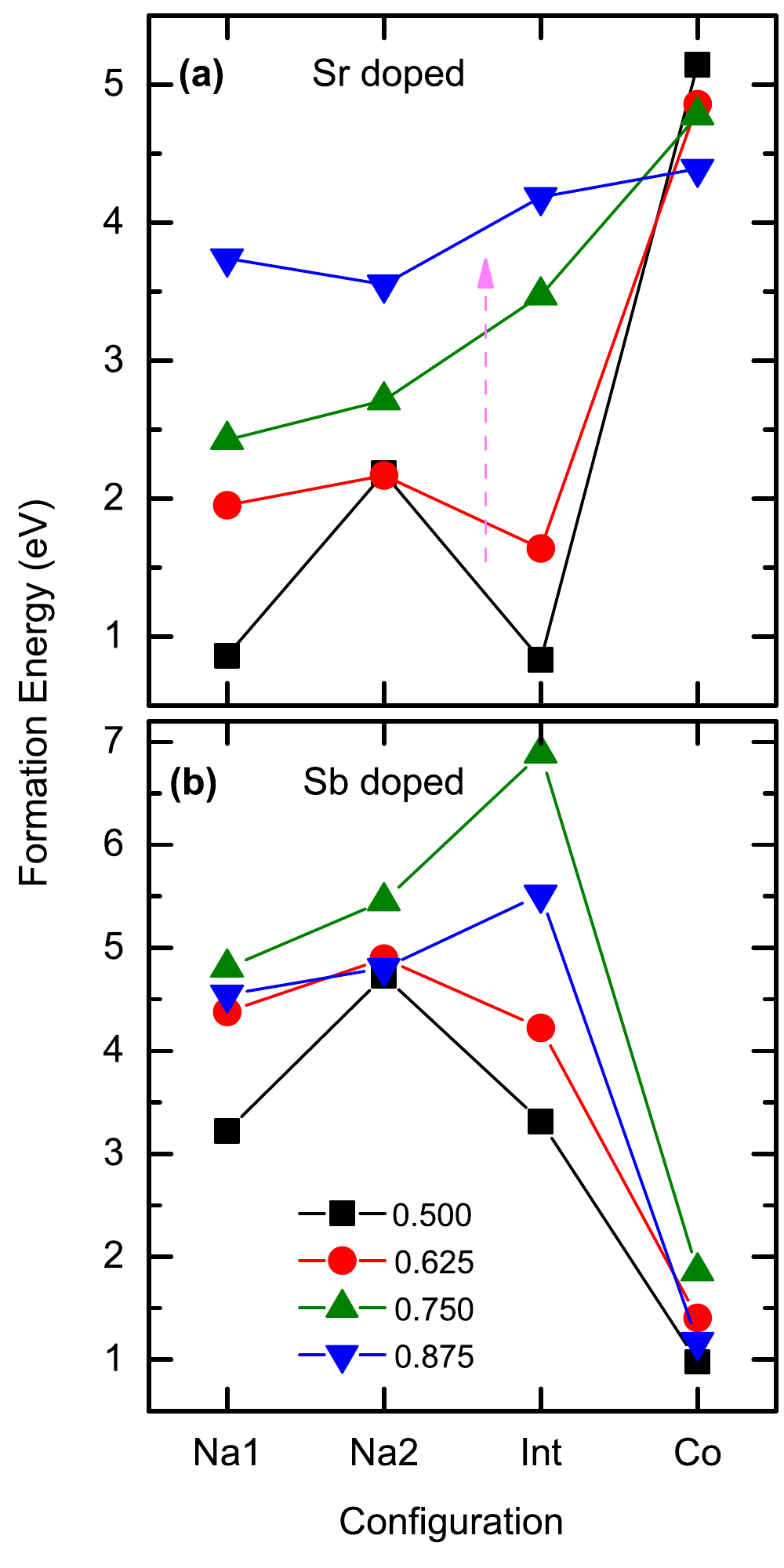}
                \caption{\label{fig:2}The formation energy of the Sr and Sb dopants in \ch{Na_{x}CoO2} are presented in (a) and (b) respectively. Na concentration ($x$) is given in the legend.}
            \end{figure}

            The formation energies of Sr dopants in \ch{Na_{x}CoO2} for the four aforementioned configurations are presented in Fig. \ref{fig:2}(a).
            In the case of $50\%$ Na concentration (\ch{Na_{0.50}CoO2:Sr}), the most stable configuration was \ch{Sr_{Int}} with an $E^f$ of $0.83 \si{.eV}$.
            The second most stable configuration, \ch{Sr_{Na 1}}, had a slightly higher $E^f$ of $0.86 \si{.eV}$.
            The third most stable configuration was \ch{Sr_{Na 2}} which had a relatively higher $E^f$ of $2.18 \si{.eV}$, while the highest $E^f$ was that of \ch{Sr_{Co}} which was $5.15 \si{.eV}$.
            For Na concentration $62.5\%$ (\ch{Na_{0.625}CoO2:Sr}), the most stable configuration was \ch{Sr_{Int}} with an $E^f$ of $1.64 \si{.eV}$ followed by \ch{Sr_{Na 1}} with an $E^f$ of $1.95 \si{.eV}$ and then by \ch{Sr_{Na 2}} that had an $E^f$ of $2.17 \si{.eV}$. The least stable configuration was \ch{Sr_{Co}} with a formation energy of $4.86 \si{.eV}$. So far the sequence of stable configurations for both Na concentrations of $50\%$ and $62.5\%$ was identical; \ch{Sr_{Int}} was the most stable configuration while \ch{Sr_{Co}} was the least stable one.

            For Na concentration of $75\%$ (\ch{Na_{0.75}CoO2:Sr}), the most stable configuration was \ch{Sr_{Na 1}} with an $E^f$ of $2.43 \si{.eV}$ followed by \ch{Sr_{Na 2}} with an $E^f$ of $2.71 \si{.eV}$ and then \ch{Sr_{Int}} with an $E^f$ of $3.47 \si{.eV}$. The least stable configuration was \ch{Sr_{Co}} with a formation energy of $4.78 \si{.eV}$. For Na concentration of $87.5\%$ (\ch{Na_{0.875}CoO2:Sr}), the most stable configuration was \ch{Sr_{Na 2}} with an $E^f$ of $3.55 \si{.eV}$ followed by \ch{Sr_{Na 1}} with an $E^f$ of $3.74 \si{.eV}$ and then by \ch{St_{Int}} with an $E^f$ of $4.19 \si{.eV}$. Once again, in a similar trend to previous Sr doped cases, the highest formation energy was that of \ch{Sr_{Co}} which was $4.39 \si{.eV}$. For higher Na concentration of $75\%$ and $87.5\%$, the Sr dopants preferred to substitute an occupied Na ion in the host lattice structure while for lower Na concentrations of $62.5\%$ and $50\%$ Sr dopants, in contrast, tended to occupy an interstitial site. Furthermore, as Na concentration increased, the difference between the formation energies of the most stable configuration and the least stable configuration (always \ch{Sr_{Co}}) decreased. This trend is marked by a pink arrow in Fig. \ref{fig:2}(a). Nonetheless, even for Na concentration of $87.5\%$, this difference in the formation energies was $0.65 \si{.eV}$ which is large enough to render the \ch{Sr_{Co}} configuration improbable for any Na concentration.

            The formation energies of Sb dopants in \ch{Na_{x}CoO2} are presented in Fig. \ref{fig:2}(b). In the case of $50\%$ Na concentration (\ch{Na_{0.50}CoO2:Sb}), the most stable configuration was \ch{Sb_{Co}} with an $E^f$ of $0.98 \si{.eV}$. The formation energies of the configurations in which Sb is incorporated in Na layer were considerably higher. \ch{Sb_{Na 1}}, \ch{Sb_{Int}} and \ch{Sb_{Na 2}} each had an $E^f$ of $3.22 \si{.eV}$, $3.32 \si{.eV}$ and $4.73 \si{.eV}$ respectively. For the Na concentration of $62.5\%$ (\ch{Na_{0.625}CoO2:Sb}), the most stable configuration was \ch{Sb_{Co}} with an $E^f$ of $1.40 \si{.eV}$ followed by —although with large margin— \ch{Sb_{Int}} with an $E^f$ of $4.22 \si{.eV}$, \ch{Sb_{Na 1}} with an $E^f$ of $4.38 \si{.eV}$ and finally \ch{Sb_{Na 2}} with $4.90 \si{.eV}$. For Na concentration of $75\%$ (\ch{Na_{0.75}CoO2:Sb}), the most stable configuration was \ch{Sb_{Co}} with an $E^f$ of $1.86 \si{.eV}$ followed by \ch{Sb_{Na 1}} with an $E^f$ of $4.81 \si{.eV}$ and \ch{Sb_{Na 2}} with an $E^f$ of $5.45 \si{.eV}$ respectively. The highest formation energy, in the case of x = $75\%$, was the one of \ch{Sb_{Int}} which was $6.89 \si{.eV}$. For Na concentration of $87.5\%$ (\ch{Na_{0.875}CoO2:Sb}), the most stable configuration was, once again, \ch{Sb_{Co}} with an $E^f$ of $1.17 \si{.eV}$ followed by \ch{Sb_{Na 1}} with an $E^f$ of $4.55 \si{.eV}$ and \ch{Sb_{Na 2}} with an $E^f$ of $4.80 \si{.eV}$. Last, in a similar trend to the case of $75\%$ Na concentration, the highest formation energy was the one of \ch{Sb_{Int}} which was $5.52 \si{.eV}$. In the case of Sb doping, in contrast to Sr doping, for all Na concentrations, the Sb dopants preferred to substitute a Co ion in the host lattice structure. Additionally, as Na concentration increased, the margin of the stability of the \ch{Sb_{Co}} configuration increased. For instance, in case $50\%$ Na concentration, the difference between the $E^f$ of \ch{Sb_{Co}} and \ch{Sb_{Na 1}} (the first and second most stable structures) was $2.24 \si{.eV}$. This difference, in the case of $87.5\%$ Na concentration, was raised to $3.37 \si{.eV}$.

            For the purpose of comparison, we note that the formation energy of oxygen vacancy ($V_O$) in \ch{Na_{x}CoO_{2-$\delta$}} is in range of $\sim 2.3 \si{.eV}$ which, consequently, results in a negligible concentration of $V_O$ \cite{Casolo2012}. For Na concentration of $50\%$, the formation energy of \ch{Sb_{Co}} (0.98 eV)  and that of \ch{St_{Int}} (0.83 eV)  are considerably lower than that of $V_O$ which indicates fabrication using equilibrium techniques such as solid state sintering may be feasible for achieving measurable dopant concentrations. However, as Na concentration increases, so is the formation energy of the dopants, especially those of Sr implying that non-equilibrium techniques such as laser bean epitaxy deposition should be used to fabricate doped \ch{Na_{x}CoO2} with higher Na concentration.

        \subsection{The Electronic Structure}
\begin{figure}
                \centering
                \includegraphics[width=0.9\columnwidth]{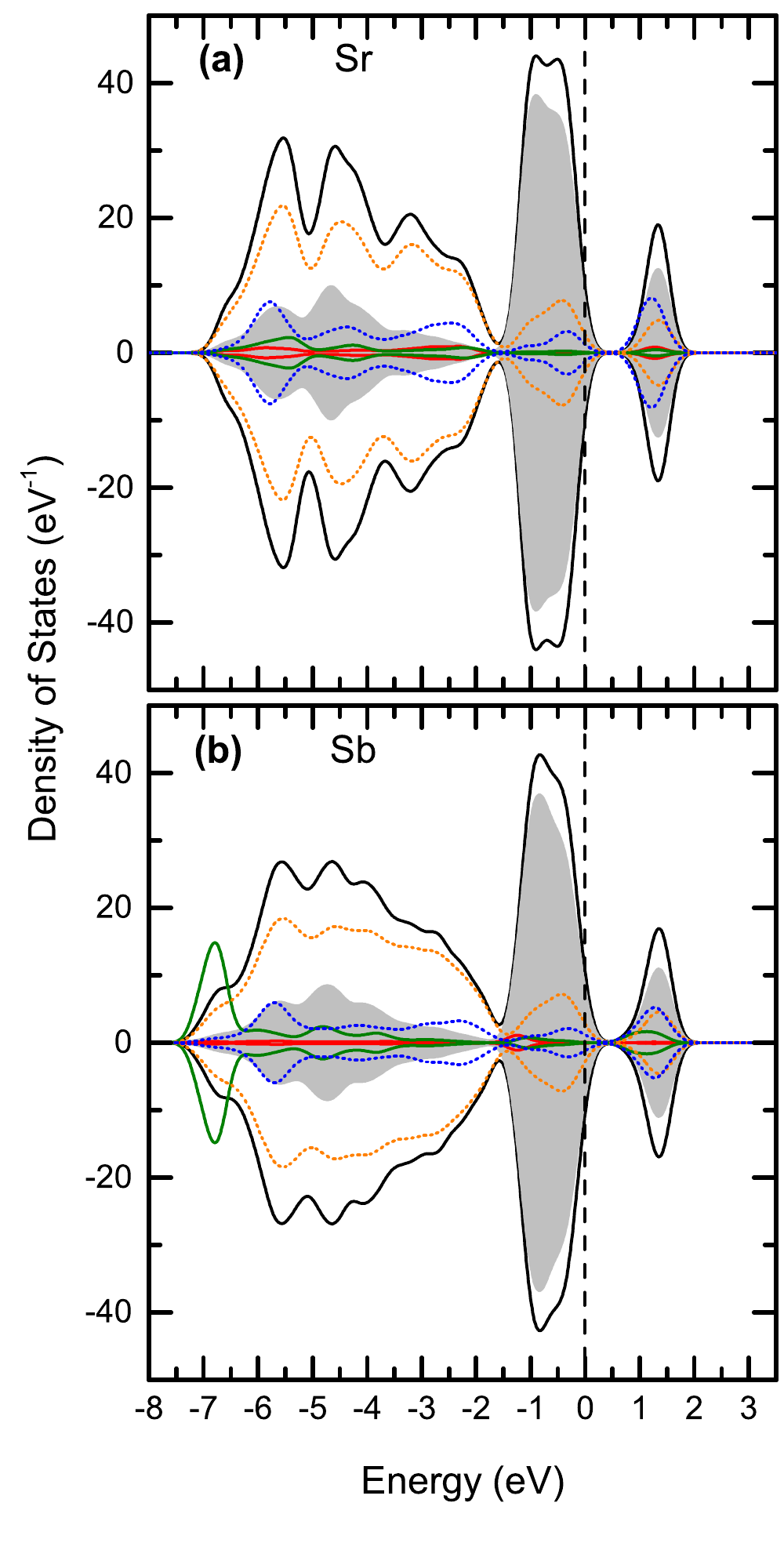}
                \caption{\label{fig:3}Total and partial density of states (DOS) of Sr and Sb doped \ch{Na_{0.75}CoO2} systems with respect to Fermi level energy ($E$\ch{_{Fermi}}). Dopants are in their respective most stable configurations, i.e. \ch{Sb_{Na}} and \ch{Sr_{Co}}. The solid black lines represent total DOS of the systems while the grey shaded areas represent Co $3d$ states. Furthermore, the solid red, solid green, dashed orange and dashed blue lines represent the dopants' $5s$, $4p$ or $5p$, O $2p$ and Na $3s$ states respectively. Sb, Sr and Na states are magnified for clarity.}
            \end{figure}

            To investigate the electronic properties, we systematically examined the total and partial density of states (P/DOS) of the Sr and Sb doped \ch{Na_{x}CoO2} systems. We found that the general features of the DOS diagrams did not strongly depend on Na concentration. Therefore, as an example, we present the DOS of \ch{Na_{0.75}CoO2:Sr_{Na 1}} and \ch{Na_{0.75}CoO2:Sb_{Co}} in Fig. \ref{fig:3}(a) and (b) respectively. As a general feature of both Sr and Sb doped systems, O $2p$ states approximately extend over an energy range of $-7  \si{.eV}$ to $-2  \si{.eV}$ while Co $3d$ states (grey shaded areas) concentrate in the region confined within $2  \si{.eV}$ below Fermi level. As a result, the hybridization between Co and O states is rather weak. This feature is very similar to the previously calculated electronic structure of \ch{Na_{x}CoO2} using different computational codes \cite{Casolo2012, Singh2000} and, as it appears, does not change by introducing Sr and Sb dopants. Fig. \ref{fig:3}(a) represents the P/DOS of the \ch{Na_{0.75}CoO2:Sr_{Na 1}} system. Both Sr $5s$ and $4p$ states are delocalized over the $-7  \si{.eV}$ to $-2  \si{.eV}$ with respect to Fermi level, hybridizing strongly with O $2p$ states. However, as Sr states diminish near Fermi level, the hybridization between Sr $5s$ and $4p$ states with Co $3d$ states becomes minimal. Furthermore, Sr DOS distribution in \ch{Na_{0.75}CoO2} appears to be very different from that of SrO in which Sr $4p$ states peak near Fermi level while Sr $5s$ states peak near the bottom of the valence band at $E  = -2\si{.eV}$ \cite{McLeod2010}. Instead, in the case of \ch{Na_{0.75}CoO2:Sr}, Sr $5s$ states strongly hybridize with Na $3s$ states as they are distributed in the same sub-bands (dashed blue lines in Fig. \ref{fig:3}(a)). As a consequent, Sr dopants are expected to behave in a similar fashion with the \ch{Na+} ions as electron donors.

            In the case of Sb doping, according to Fig. \ref{fig:3}(b), the valence band comprises a large peak of Co $t_{2g}$ states near valence band maximum while the lower region of the valence band is mainly occupied by O $2p$ states. Sb $5p$ states, to some extent, hybridize with O $2p$ states throughout the entire valence band, but they particularly peak at the bottom of the valence band near $-7  \si{.eV}$. Such electronic distribution for Sb $5p$ states is very similar to the one of \ch{Sb^V} in \ch{Sb2O5} and \ch{Sb2O4} inferring an oxidation state of V for Sb dopants in the \ch{Na_{0.75}CoO2:Sb_{Co}} system. Therefore, by substituting a \ch{Co^{3+}}, Sb dopants are expected to act, in contrast to Sr, as acceptors.

        \subsection{Alternative Doping Sites}
\begin{figure}
                \centering
                \includegraphics[width=0.9\columnwidth]{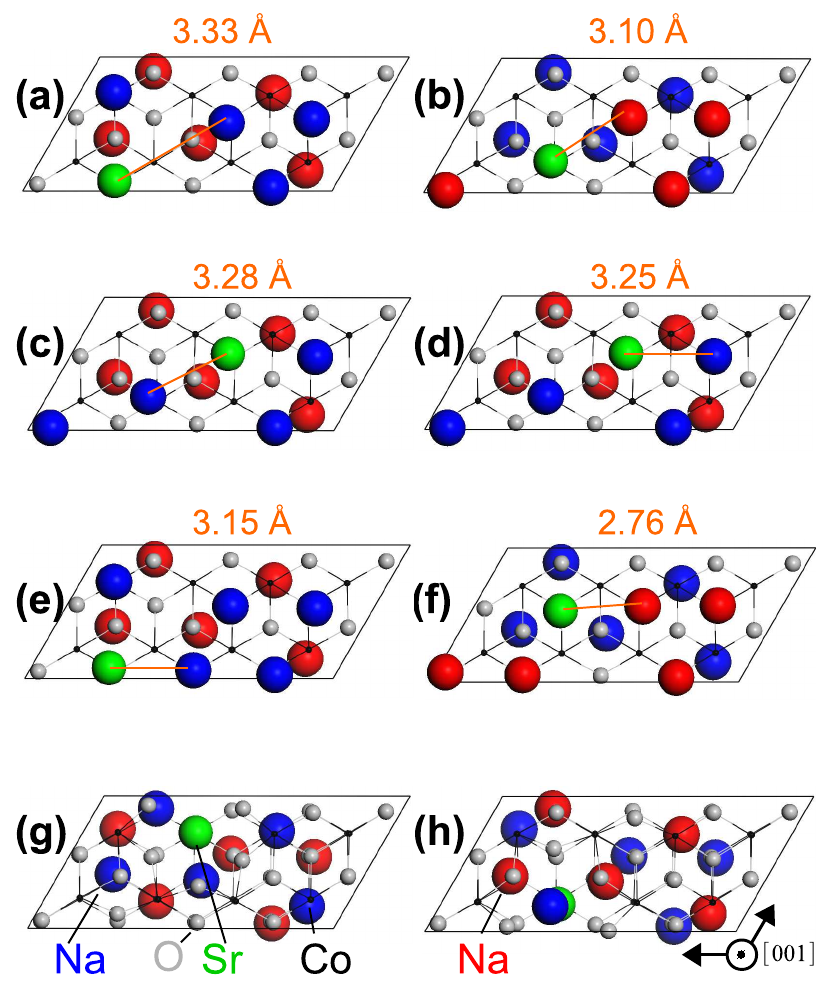}
                \caption{\label{fig:4}The original and alternative doping sites of Sr in \ch{Na_{0.625}CoO2} system are presented in the first and second column: (a) and (b) the doping sites of \ch{Sr_{Na 1}} configurations, (c) and (d) the doping sites of \ch{Sr_{Na 2}} configurations, (e) and (f) the doping sites of \ch{Sr_{Int}} configuration and finally (g) and (h) the doping sites of \ch{Sr_{Co}} configurations. For configurations (a)-(f) in which Sr dopants are located in Na layer, the distance between Sr and closest Na ion is given for each configuration.}
            \end{figure}

            The presented configurations in the previous section were fully optimized and corresponded to the lowest energy geometries, and therefore they are expected to prevail in thermodynamic equilibrium. However, non-equilibrium growth techniques such as spark plasma sintering \cite{Chen2014}, pulsed laser deposition \cite{Krochenberger2005} and reactive solid-phase epitaxy deposition \cite{Sugiura2006} methods are commonly used to fabricate doped \ch{Na_{x}CoO2} thermoelectrics. In materials fabricated by non-equilibrium methods, dopants may not occupy the thermodynamically most stable configuration. The complexity of the supercell in Na deficient systems provides many available doping sites for any given configuration of which only one is thermodynamically the most stable. In this section, we chose the Sr doped \ch{Na_{0.625}CoO2} system, as a prototype, to examine the variation of the formation energies of the four configurations when dopants were located at sites that were not of the lowest energy geometry. In the case of \ch{Sr_{Na 1}}, \ch{Sr_{Na 2}} and \ch{Sr_{Int}} configurations, we chose alternative lattice sites for which the nearest Na neighbor to the Sr dopant had shorter distance when compared to the lowest energy configuration. In of case of \ch{Sr_{Co }}, we chose a Co site that was sandwiched by Na1 ions along c direction instead of a Co site that was located in immediate vicinity of Na vacancy (lowest energy doping site). The alternative configurations are presented in Fig. \ref{fig:4} while the corresponding formation energies are presented in (a). In the case of \ch{Sr_{Na 1}} configuration, we chose an alternative Na1 site for substituting Sr that was slightly closer to the nearest neighbor Na ion within the Z = 0.5 plane. As shown in Fig. \ref{fig:4}(a) and (b), the distance between \ch{Sr_{Na 1}} and the nearest Na ion in the original and alternative configurations was $3.33\si{.\angstrom}$ and $3.10\si{.\angstrom}$ respectively. The configuration in Fig. \ref{fig:4}(a) had a formation energy of $1.95 \si{.eV}$ while the configuration in Fig. \ref{fig:4}(b) had a formation energy of $1.99 \si{.eV}$. In the case of \ch{Sr_{Na 2}} dopant, we chose a configuration that shortened the distance between \ch{Sr_{Na 2}} dopant and the nearest Na ion from $3.28\si{.\angstrom}$ to $3.25\si{.\angstrom}$ as shown in Fig. \ref{fig:4}(c) and (d). In this case, the formation energy remained constant at a value of $2.17 \si{.eV}$ for both configurations. In the case of \ch{Sr_{Int}} as in Fig. \ref{fig:4}(e) and (f), the distance between Sr dopant and the nearest Na ion within the Z = 0.5 plain was reduced from $3.15\si{.\angstrom}$ in the original configuration to $2.76\si{.\angstrom}$ in the alternative configuration which resulted in an increase in the formation energy from $4.86 \si{.eV}$ to $5.04 \si{.eV}$. In the case of \ch{Sr_{Co }}, as in Fig. \ref{fig:4}(g) and (h), the formation energy rose from $1.64 \si{.eV}$ in the original configuration to $2.64 \si{.eV}$ in the alternative configuration in which a Na shared the basal coordinate with the \ch{Sr_{Co}} dopant. By moving the Sr dopants away from their lowest energy sites to the next available neighboring site, the sequence of the stability of the configurations did not change. However, the formation energies of the Sr dopants in all configurations were slightly higher.

        \subsection{The Dependence of $E^f$ on the Choice of Functional}
            GGA methods underestimates the bandgap of most oxides \cite{Sousa2007} and therefore may cause a systematic error in the calculated formation energies. However, this error is strongly system dependent as it is a function of the calculated bandgap and the position of the dopants' electronic states with respect to the band edges. In this regard, to improve the accuracy of DFT calculations, as an \textit{ad hoc} solution, an orbital dependent Hubbard term (U) is commonly applied in order to prohibit $3d$ double-occupancy and therefore adjust the bandgap. $U$ is either tuned manually to reproduce the spectroscopic features or determined self-consistently using a linear response method \cite{Cococcioni2005}. However, it should be noted that introducing a phenomenological parameter such as $U$, renders the calculations non-first-principles. Furthermore, since dopants' formation energies strongly depend on the choice of $U$, there is always a chance that the calculated formation energies severely disagree with the experimental results \cite{Hinuma2008}. The case for the \ch{NaCoO2} systems is even more complicated as the bandgap varies with Na concentration; \ch{NaCoO2} is a band insulator with a bandgap of $\sim 1 \si{.eV}$ \cite{Boehnke2014} while Na deficient \ch{Na_{x}CoO2} exhibits metallic conduction \cite{Wang2005}. To investigate the general effect the $U$ term on the formation energies of dopants' in \ch{Na_{x}CoO2}, we calculated the formation energies using the GGA+$U$ method in the \ch{Na_{0.75}CoO2} system doped with Sb. Since many different values for $U$ have previously been proposed for the \ch{Na_{x}CoO2} system ranging from $4.0 \si{.eV}$ \cite{Wang2007} to $5.5 \si{.eV}$ \cite{Zhang2004}, as representative samples, we applied effective Hubbard values of $U\ch{_{eff}}  = 4\si{.eV}$ and $5  \si{.eV}$ within the Dudarev et al approach \cite{Dudarev1998} to Co $3d$ electrons only. The applied values for $U$\ch{_{eff}} have previously been suggested to offer a more realistic electronic description in pristine \ch{Na_{x}CoO2} \cite{Okabe2004, Lysogorskiy2012, Assadi2013}. We conducted the GGA+$U$ calculations using VASP code (as GGA+$U$ has not been implemented in \ch{DMol^3}) with projector augmented wave method (PAW) method \cite{Kresse1994, Perdew1996}. We used similar thresholds for energy and atomic forces to the one of all-electron calculations and set the energy cut off at $500 \si{.eV}$. The results are presented in (b). In the case of $U\ch{_{eff}} = 4 \si{.eV}$ , \ch{Sb_{Co}} had the lowest formation energy of $-0.12 \si{.eV}$ followed by \ch{Sb_{Na 1}} with an $E^f$ of $1.74 \si{.eV}$, then \ch{Sb_{Na 2}} with an $E^f$ of $2.39 \si{.eV}$ and finally \ch{Sb_{Int}} with an $E^f$ of $2.50 \si{.eV}$. In the case of $U\ch{_{eff}} = 5\si{.eV}$, \ch{Sb_{Co}} again had the lowest formation energy of $-2.07 \si{.eV}$ followed by \ch{Sb_{Na 2}} with an $E^f$ of $2.45 \si{.eV}$ and then \ch{Sb_{Na 1}} with an $E^f$ of $2.61 \si{.eV}$ and then \ch{Sb_{Int}} with an $E^f$ of $3.39 \si{.eV}$.

\begin{figure}
                \centering
                \includegraphics[width=0.9\columnwidth]{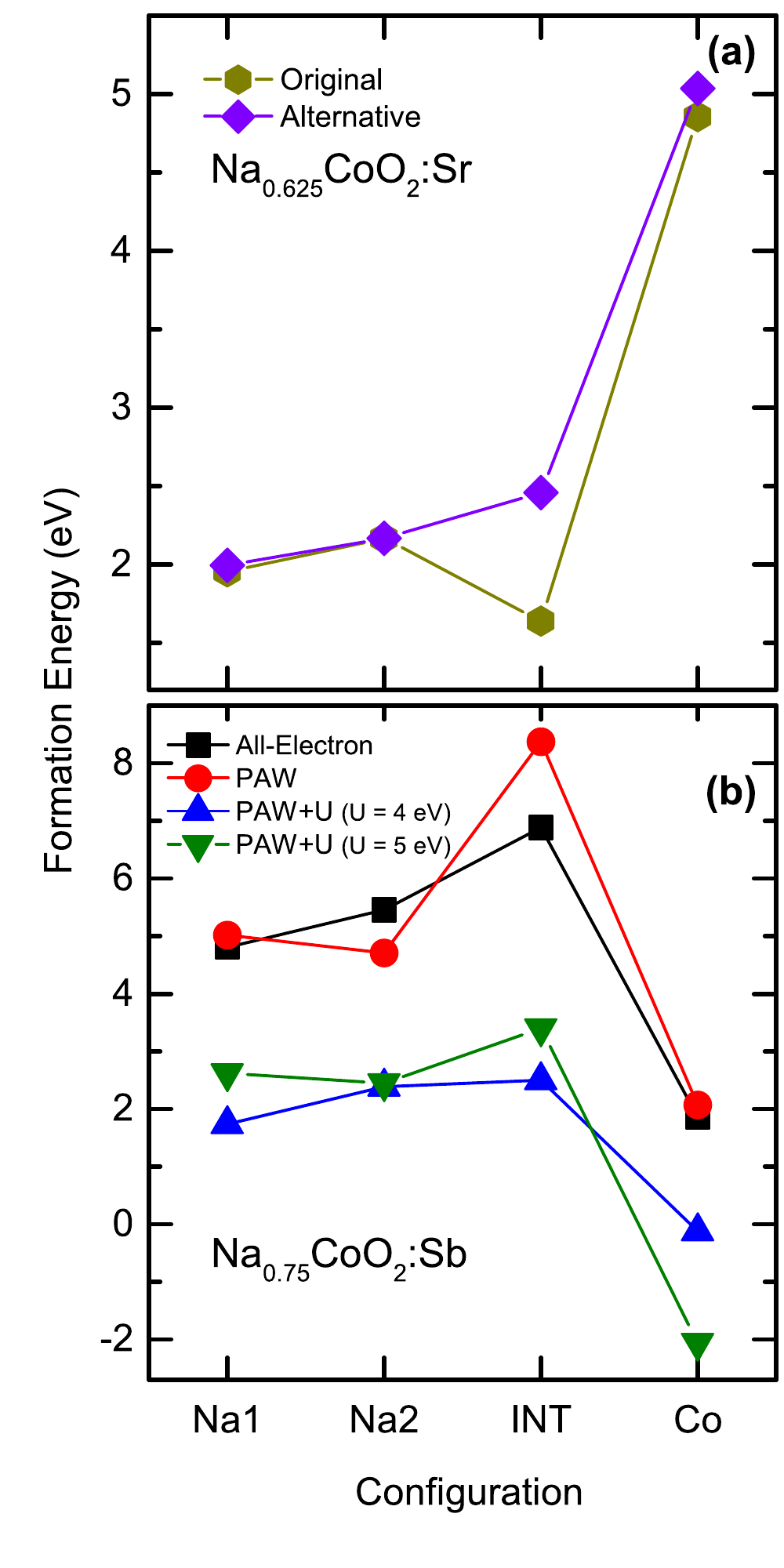}
                \caption{\label{fig:5}(a) The formation energy of the Sr dopants in \ch{Na_{0.625}CoO2} for both lowest energy doping sites (dark yellow) and alternative doping sites (purple). (b) The formation energy of Sb dopants in \ch{Na_{0.75}CoO2} based on PAW-GGA (red), PAW-GGA+$U$ (green and blue). The blue symbols represent the $E^f$ of $U\ch{_{eff}}  = 4\si{.eV}$ while the green symbols represent the formation energies of $U\ch{_{eff}}  = 5\si{.eV}$. The all-electron (\ch{DMol^3}) formation energies are also represented (red) for comparison. The minor difference in the formation energies of all-electron and PAW calculations is due to small differences in the final optimized ionic arrangements.}
            \end{figure}

\begin{figure}
                \centering
                \includegraphics[width=0.9\columnwidth]{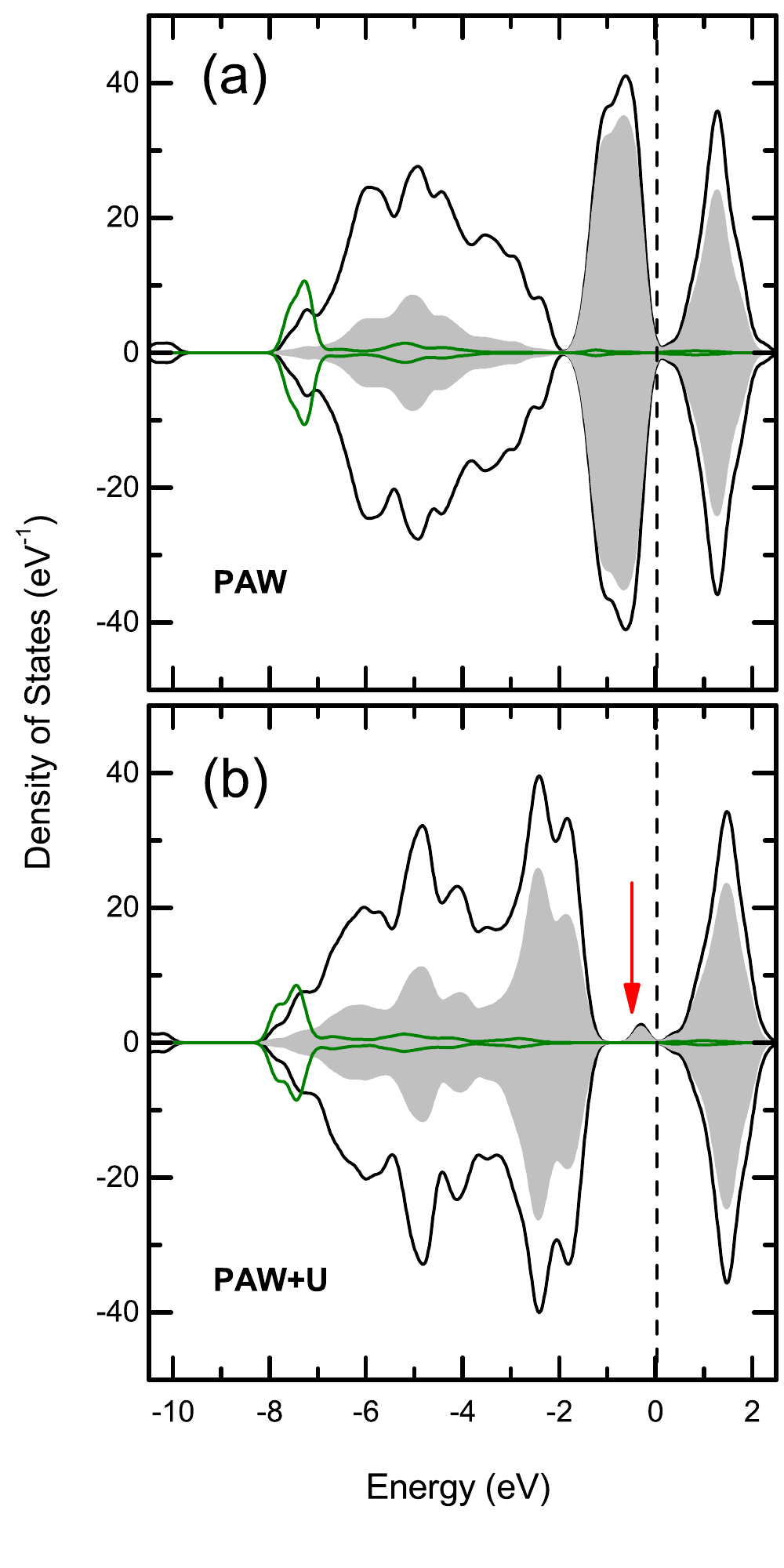}
                \caption{\label{fig:6}Total and partial density of states (DOS) of Sb doped \ch{Na_{0.75}CoO2} system calculated with PAW method with respect to $E$\ch{_{Fermi}}; (a) is the case of GGA functional while (b) is the case of GGA+$U$ functional with $U\ch{_{eff}}  = 4\si{.eV}$. The solid black lines represent total DOS of the systems while the grey shaded areas represent Co $3d$ states. The solid green lines represent the Sb $5p$ states.}
            \end{figure}

            As the trend is clear in  Fig. \ref{fig:5}(b), the sequence of the relative stability of all configurations in the Sb doped \ch{Na_{0.75}CoO2} system was not altered with the introduction of $U$\ch{_{eff}} term into the calculations. However, the absolute values of $E^f$s decreased by $\sim 2 \si{.eV}$ for \ch{Sb_{Na 1}}, \ch{Sb_{Na 2}} and \ch{Sb_{Int}} configurations for both $U\ch{_{eff}}  = 4\si{.eV}$ and $U\ch{_{eff}}  = 5\si{.eV}$. Furthermore, for the \ch{Sb_{Co}} configurations, the $E^f$ decreased by $\sim 2 \si{.eV}$ for $U\ch{_{eff}}  = 4\si{.eV}$ and by $\sim 4 \si{.eV}$ for $U\ch{_{eff}}  = 5\si{.eV}$. To identify what caused the lowered $E^f$s, the P/DOS of the \ch{Na_{0.75}CoO2:Sb_{Co}} system based on GGA and GGA+$U$ are presented in Fig. \ref{fig:6}(a) and (b) respectively. The P/DOS obtained by PAW-GGA as in Fig. \ref{fig:6}(a) is very similar to that of all-electron P/DOS of Fig. \ref{fig:3}(b). However, the introduction of the $U$\ch{_{eff}} term changed the band structure. Since the $U$\ch{_{eff}} was applied to Co $3d$, the gap between the filled $t_{2g}$ states and the unfilled $e_g$ states increased to $\sim 2 \si{.eV}$, lowering the valence band maximum by $\sim 2 \si{.eV}$ with respect to the Fermi level energy. Consequently, the hybridization between Co $3d$ states and O $2p$ states increased. However, Sb $5p$ states' position with respect to the valence band did not change, remaining at approximately $-7  \si{.eV}$ with respect to the Fermi energy. Moreover, the crystal field splitting of $t_{2g}$ states created a manifold with one peak near the Fermi level which is marked by a red arrow in Fig. \ref{fig:6}(b). The downward shift of the valence band maximum contributed to the lowering of the formation energies of all configurations regardless of the Sb's doping site while Co $3d$ higher energy $3d$ peak is the cause of a further decrease in the formation energy of \ch{Sb_{Co}}'s as Sb $5p$ states possess much lower energy.

            Although it has become clear that the absolute values $E^f$s depend on the value of $U$, we see that the sequence of the stabilization of configurations is the same regardless of the choice of $U$. At this stage, there are no experimental data on the absolute $E^f$s to clarify which value of $U$ is the most appropriate for reproducing the experimental $E^f$s. Furthermore, it has been also shown that GGA+$U$ does not perform as well as GGA in some aspects, e.g. in reproducing experimental lattice parameters for \ch{Na_{x}CoO2} when $x > 0.6$ \cite{Hinuma2008} while, in contrast, the application of $U$ term reproduces the correct \ch{Co^{3+}}/\ch{Co^{4+}} charge ordering in pristine \ch{Na_{0.75}CoO2} \cite{Meng2005}. That is because probably no unique value of $U$ simultaneously reproduces both electronic and structural details of the \ch{Na_{x}CoO2} system. Nonetheless, it is clear that regardless of the value of $U$, the most stable dopant configuration is always recognized consistently.

        \subsection{Implications on Thermoelectric $E^f$fect}
            In \ch{Na_{x}CoO2}, Co ions are in the valence state of 3+ and 4+ with six and five $3d$ shell electrons respectively. Because of the crystal field, $3d$ orbitals of the cobalt ion split into two sub-bands: a lower energy, triply degenerate band $t_{2g}$ ($d_{xy}, d_{yz}$, and $d_{xz}$) and a higher energy, doubly degenerate band $e_g$ (${{d_x}^2}{_{-y}}^2$ and ${d_z}^2$). The different possible ways of arranging the electrons in \ch{Co^{3+}} and \ch{Co^{4+}} is determined by the product of spin and orbital degeneracies and represents the entropy in the system. Now, consider an electron hopping between cobalt ions in the low spin state (schematically presented by sites 1–3 in Fig. \ref{fig:1}(b)). The band of the \ch{Co^{3+}} ion is completely filled, but the \ch{Co^{4+}} ion has an unfilled orbital and acts as a hole provider. As an electron hops from a \ch{Co^{3+}} to a \ch{Co^{4+}}, while the electrical current flows, the sites exchange a large entropy flux in the opposite direction. Therefore, in the \ch{Na_{x}CoO2} systems, an applied electric field causes a heat current of $J_Q = n \nu k_b \mbox{Ln}2$ in addition to the charge current of $J = n\nu e$ where n and $\nu$ are the hole concentration and hole's group velocity. This entropy flux is the primary origin of the unusually large Seebeck coefficient in \ch{Na_{x}CoO2}. When a dopant such as Sb substitute a Co, holes should now hop through a five center \ch{Co–O–Sb–O–Co} complex to conduct electricity and entropy. Therefore, the hole has to transfer through Co $3d$ orbitals to Sb $5p$ orbitals then back to Co $3d$ orbitals. Fig. \ref{fig:3}(b) indicates that Sb $5p$ states are located $\sim 5 \si{.eV}$ lower than the Co $t_{2g}$ states. Since resistivity is exponentially related to the energy barrier height, the electric current and entropy flux is expected to be significantly reduced by Sb dopants. As a consequence, the Seebeck coefficient in the \ch{Na_{x}CoO2:Sb} will be lowered.

            On the other hand, when a dopant such as Sr is incorporated in Na layer, it does not significantly perturb the electrical conductivity or the entropy current. This guarantees that, upon Sr doping in \ch{Na_{x}CoO2}, the system maintains its high thermopower. Additionally, the heavier and more positively charged Sr ion creates both mass and electrostatic inertia against the mobile \ch{Na+} ions at higher temperatures. As a result, by decreasing the disorder in the \ch{Na+} layer, Sr dopants will improve the overall crystallinity of the system at higher temperatures. Consequently, the mean free path of the free carriers' increases, which in turn, improves both the carriers' mobility and the system's conductivity thus raising the $ZT$. This effect has been experimentally verified for other dopants which are incorporated in the Na layer. For example, in Mg doped \ch{Na_{0.8}CoO2} system, Raman spectroscopy measurement indicated long range Na ordering at room temperatures \cite{Tsai2012}. Furthermore, neutron diffraction experiments have shown that Ca doping in Na layer creates a Na superlattice ordered over the long range at temperatures as high as $490 \si{.K}$ \cite{Porter2014}. A similar enhancement is expected in Eu doped \ch{Zn_{0.75}CoO2} for which DFT calculations predict \ch{Eu_{Na 2}} to be the most stable configuration \cite{Assadi2013}.

    \section{Conclusions}
        In this work, using all-electron density functional theory with GGA, the electronic and crystal structure of Sr and Sb doped \ch{Na_{x}CoO2} ($0.5 \leq x \leq 0.875$) was studied. We found that Sr dopants always occupy a site in the Na layer while Sb dopants always substitute a Co regardless of Na concentration. More importantly, this conclusion held when we examined the electronic correlation effects by introducing Hubbard term $U$\ch{_{eff}} into the calculations or when we considered the non-equilibrium crystallographical sites. The DOS analysis also indicated that Sr dopants were $n$-type donors while Sb dopants were $p$-type acceptors in their respective most stable configurations. By residing on the Na layer, Sr dopants, create charge and mass inertia against the liquid like Na layer, therefore, improving the crystallinity and decreasing the electrical resistivity through better carrier mobility. On the other hand, Sb dopants, by substituting a Co ion in the host matrix, reduce the electrical conductivity and the entropy flux resulting in a decreased Seebeck coefficient.
        
    \begin{acknowledgments}
        The computational facility was provided by Intersect Australia Limited. MHA acknowledges the Japan Society for Promotion of Science for his fellowship. HKY acknowledges the financial support provided by the Japan Society for the Promotion of Science for Core-to-Core Program, Computational Nano-Materials Design on Green Energy Program and a Grant-in-Aid for Materials Design through Computics, Japan Science and Technology Agency for the grant “spinodal nanotechnology for super-high-efficiency energy conversion and The Future Research Initiative Group Support Project of Osaka University on “Computational Nano-Materials Design: New Strategic Materials”.
    \end{acknowledgments}
        
    \bibliography{Paper4}{}

\end{document}